\documentclass[prd,10pt,aps,showpacs,superscriptaddress,nofootinbib,floatfix,amssymb,amsmath]{revtex4-1}
\usepackage{slashed}
\usepackage{epsfig}

\newcommand{\beqn}{\begin{eqnarray}}
\newcommand{\eeqn}{\end{eqnarray}}

\usepackage{color}

\newcommand{\revision}[1]{{#1}}
\newcommand{\revisionB}[1]{{#1}}

\begin{document}

\title{Emergent gravity in the cubic tight - binding model of Weyl semimetal in the presence of elastic deformations}


\author{Alberto Cortijo}
\affiliation{Instituto de Ciencia de Materiales de Madrid,\\
CSIC, Cantoblanco; 28049 Madrid, Spain.}


\author{M.A. Zubkov}
\email{zubkov@itep.ru}
\affiliation{Institute for Theoretical and Experimental Physics, B. Cheremushkinskaya 25, Moscow, 117259, Russia;}
\affiliation{Moscow Institute of Physics and Technology, 9, Institutskii per., Dolgoprudny, Moscow Region, 141700, Russia}
\affiliation{Far Eastern Federal University,  School of Biomedicine, 690950 Vladivostok, Russia}

\begin{abstract}
We consider the tight - binding model with cubic symmetry that may be relevant for the description of a certain class of Weyl semimetals. We take into account elastic deformations of the semimetal through the modification of hopping parameters. This modification results in the appearance of emergent gauge field and the coordinate dependent anisotropic Fermi velocity. The latter may be interpreted as emergent gravitational field.
\end{abstract}

\pacs{}

\date{\today}

\maketitle

\section{Introduction}

Recent experimental discovery of Dirac \cite{semimetal_discovery,semimetal_discovery2,semimetal_discovery3,ZrTe5,ZrTe5:2,Bi2Se3} and Weyl \cite{WeylSemimetalDiscovery} semimetals enhanced essentially the development of the interdisciplinary field of research related to the interaction between the condensed matter physics and the high energy physics. Together with $^3$He-A \cite{Volovik2003} these materials are able to serve as an arena for the experimental investigation of various effects specific for the high energy physics \cite{semimetal_effects,semimetal_effects2,semimetal_effects3,semimetal_effects4,semimetal_effects5,semimetal_effects6,semimetal_effects7,semimetal_effects8,semimetal_effects9,semimetal_effects10,semimetal_effects11,semimetal_effects12,semimetal_effects13}
because the low energy effective theory that describes fermionic quasiparticles in Dirac and Weyl semimetals has an emergent relativistic invariance \cite{Volovik2003,Parrikar2014,Vozmediano,Z2015,Chernodub:2015wxa}. However, unlike the case of the superfluid Helium, the fermions in Dirac and Weyl semimetals are charged, and experience the external magnetic and electric fields. This facilitates, in particular, the investigation of effects related to chiral anomaly \cite{ref:diffusion,ref:transport,semimetal_discovery3,ref:semimetal:2,ref:semimetal:3,ref:semimetal:4}.

Within the high energy physics there exist certain difficulties related to the inclusion into consideration of the gravitational background. In particular, there is an ambiguity in the expressions for chiral anomaly in the presence of gravity with torsion. Different methods of calculation give different expressions (see references \cite{Zanelli,Parrikar2014,Mielke,obukhov,yajima,Obukhov:1997pz,soo}, where those expressions are presented). At the same time, certain observable effects (mainly, in the condensed matter systems with emergent gravity) may be related to chiral anomaly in the presence of nontrivial geometry - for example, the appearance of Kopnin force acting on vortices in superfluid helium \cite{Volovik2003}, the appearance of the effects of anomaly in Weyl semimetals \cite{Parrikar2014,semimetal_effects2,semimetal_effects3,semimetal_effects6,semimetal_effects7,semimetal_effects8,semimetal_effects10,semimetal_effects11,semimetal_effects12,semimetal_effects13,Zyuzin:2012tv,tewary}, (in particular, chiral magnetic effect \cite{ChiralAnomalySemimetal}), etc. Therefore, the theoretical and experimental investigation of Dirac and Weyl semimetals is extremely promising since it should be able to resolve principal internal problems of the high energy physics.

In the present paper we proceed our previous theoretical investigation of Dirac and Weyl semimetals \cite{Vozmediano,Chernodub:2015wxa,Z2015}. We extend the consideration of the 2D graphene \cite{vozmediano2,vozmediano3,vozmediano4,vozmediano5,vozmediano6,VZ2013,VolovikZubkov2014,Volovik:2014kja} to the 3D Weyl semimetals. We consider the particular cubic tight - binding model \cite{WeylTightBinding}, which is able to describe qualitatively a certain class of real materials. Although this model does not describe any particular semimetal, we do not exclude, that our results may also be relevant for the quantitative study of some real materials in a certain approaximation. The regular crystal, which is described by the given model represents a Weyl semimetal with two Fermi points. Each Fermi point hosts the two  - component  Weyl fermion. The chiralities of these two Weyl fermions are different. Such materials have more reach phenomenology, than the Dirac semimetals, where each Fermi point hosts the pair of Weyl fermions of opposite chirality (i.e. the massless Dirac spinor).

We discuss the situation, when elastic deformations are present that result in the appearance of the emergent gauge field and emergent gravity. We calculate both emergent gauge field and emergent vierbein and express them through the tensor of elastic deformation.

The paper is organized as follows. In Section \ref{SectInit} we describe the general algorithm for the calculation of the emergent fields in the presence of elastic deformations that was applied earlier to graphene. In Section \ref{SectReg} we recall the description of the unperturbed tight - binding model and demonstrate, that it indeed describes Weyl semimetal. In Sect \ref{SectHop} we consider the modification of hopping parameters of the tight - binding Hamiltonian resulted from the elastic deformations. In Sect. \ref{SectElast} we calculate emergent gauge field and emergent gravitations field in the presence of elastic deformations. In Sect. \ref{SectRel} we describe the resulting low energy relativistic theory with the gauge field and the gravitational field. In Section \ref{SectCon} we end with the conclusions.

\section{How do the emergent gravity and emergent gauge field appear in Weyl semimetals}

\label{SectInit}

Similar to the previous consideration of graphene \cite{vozmediano2,vozmediano3,vozmediano4,vozmediano5,vozmediano6,Oliva,VZ2013,VolovikZubkov2014,Volovik:2014kja} we calculate the effective low energy action in the given tight - binding model of Weyl semimetal as follows:

\begin{enumerate}

\item{}
In the regular tight - binding model the positions of the Fermi points ${\bf K}^{(0)}$ are calculated as the points in momentum space, where the one  - particle Hamiltonian vanishes. The Hamiltonians are expanded around each Fermi point up to the terms linear in ${\bf p}-{\bf K}^{(0)}$.

\item{}

In the presence of elastic deformations the tight - binding model is modified.  The terms in the Hamiltonian corresponding to the jumps between the adjacent lattice sites contain modified hopping parameters \cite{WeylTightBinding,Vozmediano}. If those parameters would depend on the distance between the corresponding sites of the lattice only, then the relation between them and the tensor of elastic deformations would be similar to that of graphene \cite{vozmediano2,vozmediano3,vozmediano4,vozmediano5,vozmediano6,VZ2013,VolovikZubkov2014,Volovik:2014kja}.
Unlike graphene, however, the modification of hopping parameters in the 3D semimetals is more complicated (see \cite{WeylTightBinding,Vozmediano}). We explore the modification of the hopping parameters following \cite{Vozmediano}.

\item{}

It is assumed, that the variations of hopping parameters are weak. Therefore, we may consider space as a composition of regions such that within each region the hopping parameters do not depend on coordinates. However, within each of such regions the hopping parameters are different for different directions of links.

\item{}

Within each area, where hopping parameters may be considered as constant we calculate the position of the true Fermi point ${\bf K}$, where the one - particle Hamiltonian vanishes. This Fermi point is represented as ${\bf K}={\bf K}^{(0)} + {\bf A}$, where ${\bf A}$ is interpreted as emergent gauge field.

\item{}

Next, the one - particle Hamiltonians are expanded around each true Fermi point ${\bf K}$ in powers of ${\bf p}-{\bf K}$. The coefficients of proportionality $f^i_a$ give the tensor of anysotropic Fermi velocity (within the given region) \cite{Volovik2003}:
\begin{equation}
H = f^i_a (\hat{p}_i - K_i)
\end{equation}

\item{}

The Hamiltonian in the whole sample is written as \cite{VZ2014NPB}
\begin{equation}
H = \sigma^a f^i_a({\bf x})\circ (\hat{p}_i - K^{(0)}_i-A_i({\bf x}))=\frac{1}{2}\sigma^a \Big(f^i_a({\bf x}) (\hat{p}_i - K^{(0)}_i-A_i({\bf x}))+ (\hat{p}_i - K^{(0)}_i-A_i({\bf x}))f^i_a({\bf x})\Big)
\end{equation}
Here the values of $f^i_a({\bf x})$ and $A_i({\bf x})$ are calculated above within the regions being the small vicinities of space points $\bf x$. The composition $K^{(0)}_i-A_i({\bf x})$ is referred to as the {\it floating Fermi point}.

\item{}

Finally, the action for the fermions living near each Fermi point is written as \cite{Z2015}
\begin{equation}
S = \frac{1}{2} \int d^4x |{\bf e}| [\bar{\Psi} i  e_b^j(x) {\sigma}^b {\cal D}_j  \Psi - [{\cal D}_j\bar{\Psi}] i  e_b^j(x) {\sigma}^b  \Psi ]
\label{SHe_3sW20}
\end{equation}
Here
\begin{equation}
i{\cal D}_\mu = i\nabla_\mu +  A_\mu(x)
\end{equation}
is the covariant derivative ($A_0=0$). It corresponds to the emergent $U(1)$ gauge field $ A_\mu$. By $ e_a^j$ we denote the vierbein field, $\eta_{ab}$ is metric of Minkowski space. Internal $SO(3,1)$ indices are denoted by Latin letters $a,b,c,...$ while the space - time indices are denoted by Greek letters or Latin letters $i,j,k,...$ The inverse vierbein is denoted by $ e_i^a$ (it is assumed, that $ e^i_a  e^a_j = \delta^i_j$). The determinant of $ e^a_i$ is denoted by $|{\bf e}|$. By $\sigma^a$ we denote Pauli matrices, and imply $\sigma^0 = 1$.

The emergent vierbein is related to the tensor of anisotropic space dependent Fermi velocity as follows:
\begin{equation}
f^i_a = |{\bf e}|\, e^i_a, \quad [e^0_0]^{-1}=| {\bf e}|={\rm det}^{1/3}{\bf f}, \quad   e^{i}_0 =0, \quad e^{0}_a =0, \quad i,a = 1,2,3
\end{equation}

\end{enumerate}

\section{Regular tight - binding model}

\label{SectReg}

Let us start from the tight - binding model of lattice Wilson fermions \cite{VZWF}. It describes the $3D$ topological insulator with cubic crystal lattice and has  the following Hamiltonian:
\begin{equation}
 H_0 = \frac{1}{2}\sum_{{\bf x},j}\bar{\psi}({\bf x}+{\bf l}_j)\Big(i \, t \gamma^0 \gamma^j - r\, \gamma^0 \Big)\psi({\bf x}) + \frac{1}{2}\sum_{{\bf x}}\bar{\psi}({\bf x})(m+3\,r)\, \gamma^0  \psi({\bf x}) + (h.c.)\label{H-1}
\end{equation}
Here the sum is over the positions $\bf x$ of the 3D cubic lattice and over $j=1,2,3$. $t$ and $r$ are the hopping parameters while by $\gamma^i$  the Dirac matrices (in chiral representation) are denoted. Vectors ${\bf l}_j$ connect the nearest neighbor sites of the lattice. The given model describes Dirac fermion with mass $m$ located in the vicinity of the the point ${\bf p}=0$ in momentum space.

Extra term
\begin{equation}
 H_b = \frac{1}{2}\sum_{{\bf x}}\bar{\psi}\, b_3 \, \gamma^0 \gamma^3 \gamma^5 \psi({\bf x}) + (h.c.)
\end{equation}
transformes the given model to the model that describes Weyl semimetal. In order to demonstrate this let us consider the Hamiltonian in momentum representation:
\begin{equation}
 H = \sum_{{\bf k},j}\bar{\psi}({\bf k})\Big(\,t\,{\rm sin}({\bf k}{\bf l}_j) \, \gamma^0 \gamma^j - r\, {\rm cos}({\bf k}{\bf l}_j) \, \gamma^0 \Big)\psi({\bf k}) + \sum_{{\bf k}}\bar{\psi}({\bf k})\Big((m+3\,r)\, \gamma^0 +  b_3 \, \gamma^0 \gamma^3 \gamma^5  \Big)\psi({\bf k})\label{H0}
\end{equation}
Here ${\bf k}$ is momentum of the quasiparticle.
Next, let us define
\begin{equation}
p_i = \frac{{\rm sin}\,(k_i \,a)}{a}
\end{equation}
 (where $a = |{\bf l}_i|$ is the lattice spacing), and
\begin{equation}
F = m + r \, \sum_{i=1,2,3} (1 - {\rm cos}\, {\bf k}{\bf l}_i)
\end{equation}
and apply the transformation $\psi \rightarrow (\gamma^0-\gamma^0\gamma^3)\psi$. This gives the transformed one - particle Hamiltonian (${\bf p}_\bot=(p_1,p_2)$):
\begin{equation}
 H = \left(\begin{array}{cc}v{\bf p}_\bot \sigma + (F-b_3)\sigma^3 & - v p_3\\
 -v p_3 & -v {\bf p}_\bot \sigma - (F+b_3)\sigma^3\end{array} \right)\label{H1}
\end{equation}
Here we take into account that the vectors ${\bf l}_j$ form the triad of the cubic lattice.
We introduce the dimensionless parameter
\begin{equation}
v = t\, a,
\end{equation}
where $a$ is the lattice spacing. It will be seen below, that this parameter has the meaning of the Fermi velocity in the $xy$ plane. Dirac spinor has the form of the pair of two - component spinors
\begin{equation}
\psi = \left(\begin{array}{c} \psi_{1}\\ \psi_{2} \end{array}\right)
\end{equation}
Let us denote energy by ${\cal E}(p)$. Then we have:
\begin{equation}
\psi_{2} = - \frac{v p_3}{{\cal E}(p)+v {\bf p}_\bot \sigma + (F+b_3)\sigma^3}  \psi_{1}  \approx - \frac{v p_3}{b_3+m} \sigma^3 \psi_{1}
\end{equation}
For small energies (close to the Fermi points) we arrive at
\begin{equation}
\psi_{2} \approx - \frac{v p_3}{v {\bf p}_\bot \sigma + (F+b_3)\sigma^3}  \psi_{1}
\end{equation}
and the Hamiltonian for the reduced two - component spinors receives the form
\begin{eqnarray}
 H_{reduced} & = & v {\bf p}_\bot \sigma + {\sigma^3}{(F-b_3)} + \frac{v {\bf p}_\bot \sigma + (F+b_3)\sigma^3}{v^2 {\bf p}^2_\bot  + (F+b_3)^2} v^2 p_3^2\nonumber\\
 & = & v {\bf p}_\bot \sigma \Big(1+\frac{p_3^2}{{\bf p}^2_\bot  + \frac{(F+b_3)^2}{v^2}} \Big)+ {\sigma^3}\Big({(F-b_3)} + \frac{(F+b_3)}{v^2 {\bf p}^2_\bot  + (F+b_3)^2} v^2 p_3^2\Big)
\end{eqnarray}
For $\sqrt{\frac{b_3^2-m^2 }{v^2 + \frac{a^2}{2}r}}\ll \frac{\pi}{a}$ it receives the especially simple form (up to the terms quadratic in $p$):
\begin{equation}
 H_{reduced} = v {\bf p}_\bot \sigma + \frac{\sigma^3}{m+b_3} \Big[(m+b_3)\Big(m + r \, \sum_{i=1,2,3} (1 - {\rm cos}\, {\bf k}{\bf l}_i) - b_3\Big) + v^2 p_3^2 \Big]
\end{equation}
There are two Dirac points
\begin{equation}
\pm {\bf K}^{(0)} = \Big(0, 0, \pm K^{(0)}_3\Big)
\end{equation}
with
\begin{equation}
K^{(0)}_3 \approx  \sqrt{\frac{b_3^2-m^2 }{v^2 + \frac{a^2}{2}r(m+b_3)}}
\end{equation}
Let us denote
\begin{equation}
\lambda = \sqrt{1 + \frac{a^2}{2 v^2}r(m+b_3)} =  \sqrt{1 + \frac{r}{2 t^2}(m+b_3)}
\end{equation}
\revision{Notice, that in the original model of lattice Wilson fermions $v = t\, a \sim r\, a \sim 1$, while $m$ has the meaning of the physical mass. In this case the value of $\lambda$ would satisfy $\lambda \approx 1$ in the limit $a\rightarrow 0$. In our case due to the smalness of $v \sim 1/200$ the value of $\lambda$ may, in principle, differ from unity. However, we will see below, that rather natural assumptions about the components of the anisotropic Fermi velocity tensor will give $\lambda \approx 1$.}

The Hamiltonians near these Fermi points have the form:
\begin{equation}
 H_{+} = v {\bf p}_\bot \sigma + {v} \lambda (p_3-K^{(0)}_3) \sigma^3 2\sqrt{\frac{b_3-m}{b_3+m}}
\end{equation}
and
\begin{equation}
 H_{-} = v {\bf p}_\bot \sigma - {v} \lambda (p_3+K^{(0)}_3) \sigma^3 2\sqrt{\frac{b_3-m}{b_3+m}} = -\sigma^3 \Big( v {\bf p}_\bot \sigma + {v}\lambda (p_3+K^{(0)}_3) \sigma^3 2\sqrt{\frac{b_3-m}{b_3+m}}\Big) \sigma^3
\end{equation}
After the transformation $\psi_-\rightarrow \sigma^3 \psi_-$ of the two - component spinor at  $-{\bf K}^{(0)}$ we arrive at the model with the right - handed fermion at $+{\bf K}^{(0)}$ and the left - handed fermion at $-{\bf K}^{(0)}$.

The Fermi velocity is anysotropic and has the form
\begin{equation}
{\bf f}^{(0)} = v \left(\begin{array}{ccc}1&0&0\\
0&1&0\\
0&0& 2\lambda \sqrt{\frac{b_3-m}{b_3+m}}  \end{array}\right)\label{f0}
\end{equation}
Notice, that the given low energy theory may be applied if $|{\bf K}^{(0)}|a \ll \pi$, so that the two Weyl points are close enough to each other and we may indeed use Eq. (\ref{H1}) instead of Eq. (\ref{H0}).

\revision{In the following we consider for simplicity the case, when
$$r \, a  \lesssim t\, a = v \ll 1.$$ As it was mentioned, we assume, that the Fermi point value is much smaller, than $\pi/a$, i.e.
$$K^{(0)}_3 = \frac{\sqrt{b_3^2-m^2}}{v} = \gamma \pi/a$$ with $\gamma \ll 1$. Then $$f^3_3 =  2 v  \lambda \sqrt{\frac{b_3-m}{b_3+m}} = 2 v \,\lambda  \frac{v\, \gamma \pi}{a\,(b_3+m)}.$$ In the typical Weyl semimetals the components of the anisotropic Fermi velocity tensor are of the same order. Therefore, $a\,(b_3+m) \sim v\, \gamma \pi \lambda$, which is consistent with the value of fermion mass $m \sim v\, \gamma \, \lambda \frac{\pi}{a} $ (of the original topological insulator Eq. (\ref{H-1})) much smaller than $K^{(0)}_3 \ll \pi/a$. The value of $\lambda$ satisfies
\begin{equation}
|\lambda^2-1| =  |\frac{r}{2 t^2}(m+b_3)| \lesssim  |\frac{1}{2 t}(m+b_3)| = |\frac{a}{2 v}(m+b_3)| \sim \gamma \pi \lambda \ll \lambda
\end{equation}
From here it follows, that $\lambda \approx 1$.
}

\section{Modification of hopping parameters resulted from elastic deformations}

\label{SectHop}

In the presence of elastic deformations \cite{ref:LL} the hopping parameters depend on direction and on the position in space. First let us consider the simplest model that relates hopping parameters with the tensor of elastic deformations. In this model the hopping parameter corresponding to the jump between the two sites ${\bf x}$ and ${\bf x}+{\bf l}_j$ depends only on the real distance between these two sites given by ${r}({\bf x},{\bf l}_j) = |{\bf u}({\bf x}+{\bf l}_j)-{\bf u}({\bf x})|$, where $\bf u$ is the displacement vector. Therefore, we substitute the hopping parameter standing at the link $({\bf x},j)$ (summation over $k$ and $m$ is assumed):
\begin{equation}
t \rightarrow t (1 - \beta_t\, l^k_j l^m_j u_{km}) = t (1-\beta_t u_{jj})
\end{equation}
and
\begin{equation}
r \rightarrow r (1 - \beta_r\, l^k_j l^m_j u_{km}) = r (1-\beta_r u_{jj})
\end{equation}
Here
\begin{equation}
u_{km} = \frac{1}{2}(\partial_k u_m + \partial_m u_k)
\end{equation}
is the linearized deformation tensor.

Now let us consider the following complication. Let us assume, that the parameter $t$ recieves extra correction due to the non - diagonal elements of the deformation tensor \cite{WeylTightBinding}:
\begin{equation}
t \gamma^j \rightarrow t \gamma^j (1 - \beta_t\, l^k_j l^m_j u_{km}) + t \beta^\prime_t \sum_{n\ne j} l^k_j l^m_n u_{km} \gamma^n = t \gamma^j (1-\beta_t u_{jj}) + t \beta^\prime_t \sum_{n\ne j} u_{jn} \gamma^n
\end{equation}

\section{Tight - binding model in the presence of elastic deformations}

\label{SectElast}

\subsection{Appearance of emergent gauge field}

The modified tight - binding model corresponds to the Hamiltonian
 \begin{eqnarray}
 H &=& \frac{1}{2}\sum_{{\bf x},j}\bar{\psi}({\bf x}+{\bf l}_j)\Big(i \, t (1-\beta_t u_{jj}) \gamma^0 \gamma^j + i  t \beta^\prime_t \sum_{n\ne j} u_{jn} \gamma^0 \gamma^n - r (1-\beta_r u_{jj})\, \gamma^0 \Big)\psi({\bf x})\nonumber\\&& + \frac{1}{2}\sum_{{\bf x}}\bar{\psi}({\bf x})\Big((m+3\,r)\, \gamma^0 +  b_3 \, \gamma^0 \gamma^3 \gamma^5  \Big)  \psi({\bf x}) + (h.c.)
\end{eqnarray}
Let us denote the trace of the deformation tensor by
\begin{equation}
u \equiv \sum_{j=1,2,3}u_{jj}
\end{equation}
Repeating the above steps we arrive at
the one - particle Hamiltonian
\begin{equation}
 H = \left(\begin{array}{cc}\begin{array}{c}v  \sum_{i=1,2}\sigma^i (\hat{p}_i  (1-\beta_t u_{ii}) +   \beta^\prime_t \sum_{n\ne i} u_{in} \hat{p}_n )\\ + \Big(m + r \, \sum_{i=1,2,3} \Big[1 - (1 -  \beta_r  u_{ii}){\rm cos}\, {\bf k}{\bf l}_i\Big] -b_3\Big)\sigma^3 \end{array}& - v \hat{p}_3 (1-\beta_t u_{33}) - v \beta^\prime_t \sum_{n\ne 3} u_{in} \hat{p}_n \\
 -v \hat{p}_3 (1-\beta_t u_{33})-v \beta^\prime_t \sum_{n\ne 3} u_{in} \hat{p}_n  & \begin{array}{c}-v \sum_{i=1,2}  \sigma^i (\hat{p}_i(1-\beta_t u_{ii}) +   \beta^\prime_t \sum_{n\ne i} u_{in} \hat{p}_n ) \\  -  \Big(m + r \, \sum_{i=1,2,3} \Big[1 - (1 -  \beta_r  u_{ii}){\rm cos}\, {\bf k}{\bf l}_i\Big] +b_3\Big)\sigma^3\end{array}\end{array} \right)\label{H2}
\end{equation}
Here the product of momentum operator $\hat{p}_k$ and the coordinate dependent function $u_{ij}({\bf x})$ is defined as $\frac{1}{2} (\hat{p}_k u_{ij}({\bf x}) + u_{ij}({\bf x}) \hat{p}_k)$.
Eq. (\ref{H2}) has the form of Eq. (\ref{H1}) with the substitution
\begin{eqnarray}
\hat{p}_i & \rightarrow & \hat{P}_i = \hat{p}_i  (1-\beta_t u_{ii}) +   \beta^\prime_t \sum_{n\ne i} u_{in} \hat{p}_n\nonumber\\
m &\rightarrow & F = m + r \, \sum_{i=1,2,3} \Big[1 - (1 -  \beta_r  u_{ii}){\rm cos}\, {\bf k}{\bf l}_i\Big]
\nonumber\\
&\approx & M + r \, \sum_{i=1,2,3} \Big[(1 -  \beta_r  u_{ii})(1-{\rm cos}\, {\bf k}{\bf l}_i)\Big]
\nonumber\\
&& M = m + r \,  \beta_r  u
\end{eqnarray}

The reduced Hamiltonian receives the form
\begin{eqnarray}
H_{reduced} &=& v \sum_{i=1,2} P_i \sigma^i + \frac{\sigma^3}{M+b_3} \Big[(M+b_3)\Big(M + r \, \sum_{i=1,2,3} (1 -  \beta_r  u_{ii})(1 - {\rm cos}\, {\bf k}{\bf l}_i) - b_3\Big) + v^2 P_3^2 \Big]\nonumber\\
&\approx & v \sum_{i=1,2} P_i \sigma^i + \frac{\sigma^3}{M+b_3} \Big[(M+b_3)\Big(M - b_3\Big) + v^2 P_3^2 \Big]
\end{eqnarray}
\revision{Here we neglected the terms proportional to $\frac{a^2}{2 v^2}r(M+b_3) $ following the unperturbed model with $\lambda \approx 1$. Recall that we consider the case, when
$$r \, a  \lesssim t\, a = v \ll 1$$
Besides, we assume, that the Fermi point value is much smaller, than $\pi/a$, i.e.
$$K^{(0)}_3 = \frac{\sqrt{b_3^2-m^2}}{v} = \gamma \pi/a, \quad \gamma \ll 1$$ The third component of the unperturbed Fermi velocity is given by $$f^{(0)3}_3 =  2 v  \lambda \sqrt{\frac{b_3-m}{b_3+m}} = 2 v \,\lambda  \frac{v\, \gamma \pi}{a\,(b_3+m)}$$
The requirement, that this component is of the same order as $f^{(0)1}_1 = f^{(0)2}_2 = v$ gives
 $$a\,(b_3+m) \sim v\, \gamma \pi \lambda$$ Therefore
\begin{equation}
|\frac{a^2}{2 v^2}r(M+b_3)|  = |\frac{a^2}{2 v^2}r(m+r \beta_r u + b_3)| \sim  |\frac{r}{2 t^2}(m+b_3) + \frac{\beta_r}{2} u| \lesssim  |\frac{1}{2 t}(m+b_3)| = |\frac{a}{2 v}(m+b_3)| \sim \gamma \pi \lambda \ll \lambda \approx 1
\end{equation}
This sequence of relations demonstrates, that we may indeed neglect $\frac{a^2}{2 v^2}r(M+b_3)$.}
The position of the floating Fermi point corresponds to the following values of $P$:
\begin{equation}
{\bf P} = \Big(0, 0, \pm \frac{\sqrt{b_3^2-M^2}}{v}\Big)
\end{equation}
The corresponding values of momentum are related to these values as follows:
\begin{equation}
 \left(\begin{array}{ccc}1-\beta_t u_{11}&\beta^\prime_t u_{12}&\beta^\prime_t u_{13}\\
\beta^\prime_t u_{11}&1-\beta_t u_{22}&\beta^\prime_t u_{23}\\
\beta^\prime_t u_{31}&\beta^\prime_t u_{32}& 1-\beta_t u_{33}  \end{array}\right)\left(\begin{array}{c}p_1 \\ p_2 \\ p_3 \end{array}\right) = \left(\begin{array}{c}0 \\ 0 \\ \pm \frac{\sqrt{b_3^2-(m+r \beta_r  u)^2}}{v } \end{array}\right)
\end{equation}
The solution is (up to the terms linear in $u$):
\begin{equation}
\left(\begin{array}{c}p_1 \\ p_2 \\ p_3 \end{array}\right) \approx \pm \left(\begin{array}{ccc}1+\beta_t u_{11}&-\beta^\prime_t u_{12}&-\beta^\prime_t u_{13}\\
-\beta^\prime_t u_{11}&1+\beta_t u_{22}&-\beta^\prime_t u_{23}\\
-\beta^\prime_t u_{31}&-\beta^\prime_t u_{32}& 1+\beta_t u_{33}  \end{array}\right)  \left(\begin{array}{c}0 \\ 0 \\ \frac{\sqrt{b_3^2-m^2}}{v}(1-\frac{m r \beta_r u}{b_3^2-m^2}) \end{array}\right)
\end{equation}
This gives the position of the floating Fermi point
\begin{equation}
\pm {\bf K} = \pm \Big(-\beta^\prime_t u_{13}, -\beta^\prime_t u_{23}, 1+ \beta_t u_{33} - \frac{m r \beta_r u}{b_3^2-m^2}\Big)\frac{\sqrt{b_3^2-m^2}}{v})=\pm ({\bf K}^{(0)}+ {\bf A}),
\end{equation}
where the emergent gauge field has the following components
\begin{eqnarray}
A_1 &= & -{\beta^\prime_t} u_{13}\frac{\sqrt{b_3^2-m^2}}{v}\nonumber\\
A_2 &= & -{\beta^\prime_t} u_{23}\frac{\sqrt{b_3^2-m^2}}{v}\nonumber\\
{A}_3 &=& \Big(\beta_t u_{33} - \frac{m r \beta_r u}{b_3^2-m^2}\Big)\frac{\sqrt{b_3^2-m^2}}{v}\label{A3}
\end{eqnarray}

\revisionB{Notice, that $\frac{m r }{b_3^2-m^2}\sim \frac{a^2 m r }{v^2 \gamma^2 \pi^2}\sim \frac{a r    \lambda}{v \gamma \pi} \sim \frac{ a r  }{ v \gamma \pi}$. This is natural to suppose, that in the given semimetal all nonzero components of the emergent gauge field are of the same order. This requirement gives the natural constraint on the hopping parameter $r$: $$r \sim \frac{v \gamma \pi}{a}\sim m$$ In this case both terms in the last row of Eq. (\ref{A3}) are of the same order.}

\subsection{Appearance of space depending anisotropic Fermi velocity}

The linearized Hamiltonians near the floating Fermi points have the form:
\begin{equation}
 H_{+} = v \sum_{i=1,2} P_i \sigma^i + v \Big(P_3-\frac{\sqrt{b_3^2-M^2}}{v}\Big) \sigma^3 2\sqrt{\frac{b_3-M}{b_3+M}}
\end{equation}
and
\begin{equation}
 H_{-} = -\sigma^3 \Big( v \sum_{i=1,2} P_i \sigma^i + v \Big(P_3+\frac{\sqrt{b_3^2-M^2}}{v}\Big) \sigma^3 2\sqrt{\frac{b_3-M}{b_3+M}}\Big) \sigma^3
\end{equation}
Therefore, tensor of anisotropic Fermi velocity receives the form
\begin{equation}
{\bf f} \approx  v \, \left(\begin{array}{ccc}1& 0 & 0\\
0 &1& 0\\
0 & 0 &  (1-\frac{b_3 r \beta_r u}{b_3^2-m^2})\,2\sqrt{\frac{b_3-m}{b_3+m}} \end{array}\right)\left(\begin{array}{ccc}1-\beta_t u_{11}&\beta^\prime_t u_{12}&\beta^\prime_t u_{13}\\
\beta^\prime_t u_{11}&1-\beta_t u_{22}&\beta^\prime_t u_{23}\\
\beta^\prime_t u_{31}&\beta^\prime_t u_{32}& 1-\beta_t u_{33}  \end{array}\right)
\end{equation}
that is
\begin{equation}
{\bf f} \approx v \left(\begin{array}{ccc}1-\beta_t u_{11}&\beta^\prime_t u_{12}&\beta^\prime_t u_{13}\\
\beta^\prime_t u_{11}&1-\beta_t u_{22}&\beta^\prime_t u_{23}\\
\beta^\prime_t u_{31}&\beta^\prime_t u_{32}& (1-\beta_t u_{33}-\frac{b_3 r \beta_r u}{b_3^2-m^2})\,2\sqrt{\frac{b_3-m}{b_3+m}}  \end{array}\right)
\end{equation}
We may also rewrite this expression in the following way:
\begin{equation}
f^i_a \approx  f^{(0),j}_a \Big(\delta^i_j + \beta_t^\prime u_{ij} - \delta^i_{3}\delta_j^3\,\frac{b_3 r \beta_r u}{b_3^2-m^2} - (\beta_t^\prime + \beta_t)\,\sum_b \delta^i_{b}\delta_{jb} u_{bb}  \Big),
\end{equation}
where the unperturbed Fermi velocity tensor $f^{(0),i}_a$ is given by Eq. (\ref{f0}).

The Hamiltonians at the (floating) Fermi points have the form:
\begin{equation}
 H_{+} = f^i_a \sigma^a (p_i-K_i)
\end{equation}
and
\begin{equation}
 H_{-} =  -\sigma^3 \Big(f^i_a \sigma^a (p_i+K_i) \Big) \sigma^3
\end{equation}

\section{Relativistic fermions in Weyl Semimetal in the presence of elastic deformations}

\label{SectRel}

Near each of the two Fermi points there is the left-handed/right - handed Weyl fermion.  In the presence of elastic deformations the action for the right - handed Weyl fermion (incident at the Fermi point $+{\bf K}$) has the form \cite{Z2015}:
\begin{equation}
S_R = \frac{1}{2} \int d^4x |{\bf e}| [\bar{\Psi} i  e_b^j(x) {\sigma}^b {\cal D}_j  \Psi - [{\cal D}_j\bar{\Psi}] i  e_b^j(x) {\sigma}^b  \Psi ]
\label{SHe_3sW2}
\end{equation}
Here
\begin{equation}
i{\cal D}_\mu = i\nabla_\mu +  A_\mu(x)
\end{equation}
is the covariant derivative. It corresponds to the $U(1)$ gauge field $ A_\mu$ with the only nonzero component given by Eq. (\ref{A3}).
The left - handed fermion belongs to the point $-{\bf K}$ and has the action
\begin{equation}
S_L  =  \frac{1}{2} \int d^4x |{\bf e}| [\bar{\Psi} i  e_b^j(x) \bar{\sigma}^b {\cal D}_j  \Psi - [{\cal D}_j\bar{\Psi}] i  e_b^j(x) \bar{\sigma}^b  \Psi ]
\label{SHe_3sW6}
\end{equation}
Here $\bar{\sigma}^0 = 1$,  $\bar{\sigma}^a = - \sigma^a$ for $a=1,2,3$ while
\begin{equation}
i{\cal D}_\mu = i\partial_\mu -  A_\mu(x)
\end{equation}
The vierbein in the absence of elastic deformations is given by
\begin{equation}
[e^{(0)0}_0]^{-1}=| {\bf e}^{(0)}|=v_F, \quad  e^{(0)i}_a =  \hat{f}^i_a,\quad  e^{(0)i}_0 =0, \quad e^{(0)0}_a =0,\quad v_F = 2^{1/3}\, v \, \Big(\frac{b_3-m}{b_3+m}\Big)^{1/6}
 \label{Connection00}
\end{equation}
where $a,i,j,k=1,2,3$, while $f^i_a=v_F \hat{f}^i_a$ has the meaning of the unperturbed anisotropic Fermi velocity
with the the $3\times 3$ matrix
\begin{equation}
\hat{f} = {\rm diag}(\nu^{-1/3},\nu^{-1/3},\nu^{2/3}), \quad \nu = 2\sqrt{\frac{b_3-m}{b_3+m}}
\end{equation}

In the presence of elastic deformations the emergent vielbein is expressed up to the terms linear in displacement vector as follows:
\begin{eqnarray}
  e_a^{i} & = & \hat{f}_a^{i}\Big(1+ (\beta_t + \beta_r \frac{r b_3}{b_3^2-m^2})\frac{1}{3} u\Big)  + \hat{f}^{k^\prime}_a U_{k^\prime}^k \nonumber\\  e_0^{i} & = & 0, \quad  e^0_a = 0\nonumber\\
  e_0^0 & = & \frac{1}{v_F}\Big(1+ (\beta_t + \beta_r \frac{r b_3}{b_3^2-m^2})\frac{1}{3} u\Big)
 \nonumber\\
 |{\bf e}| & = &  v_F\Big(1- (\beta_t + \beta_r \frac{r b_3}{b_3^2-m^2})\frac{1}{3} u\Big)\nonumber\\ &&  a,i, j, k,n = 1,2,3\label{eg}
\end{eqnarray}
Here
\begin{equation}
{\bf U} =  \left(\begin{array}{ccc}-\beta_t u_{11}&\beta^\prime_t u_{12}&\beta^\prime_t u_{13}\\
\beta^\prime_t u_{11}&-\beta_t u_{22}&\beta^\prime_t u_{23}\\
\beta^\prime_t u_{31}&\beta^\prime_t u_{32}& (-\beta_t u_{33}-\frac{b_3 r \beta_r u}{b_3^2-m^2})  \end{array}\right)
\end{equation}
In another notations we have:
\begin{equation}
U^i_j = \Big(\beta_t^\prime u_{ij} - \delta^i_{3}\delta_j^3\,\frac{b_3 r \beta_r u}{b_3^2-m^2} - (\beta_t^\prime + \beta_t)\,\sum_b \delta^i_{b}\delta_{jb} u_{bb}  \Big)
\end{equation}
The emergent gauge field is given by
\begin{eqnarray}
 A_1 &= & -{\beta^\prime_t} u_{13}\frac{\sqrt{b_3^2-m^2}}{v}\nonumber\\
A_2 &= & -{\beta^\prime_t} u_{23}\frac{\sqrt{b_3^2-m^2}}{v}\nonumber\\
{A}_3 &=& \Big(\beta_t u_{33} - \frac{m r \beta_r u}{b_3^2-m^2}\Big)\frac{\sqrt{b_3^2-m^2}}{v}\nonumber\\ A_0^{} &=& 0
 \label{DiracPosition1_}
\end{eqnarray}
(Recall, that $u=\sum_i u_{ii}$.)
The analogy to graphene prompts that the  values of $\beta_t$, $\beta_t^\prime$, and $\beta_r$ similar to the Gruneisen parameter are of the order of unity.

In the particular case, when $\beta_t = - \beta_t^\prime$, $\beta_r = 0$ the expressions for the emergent fields are especially simple:
\begin{eqnarray}
 A_i & = & {\beta_t} u_{i3}\frac{\sqrt{b_3^2-m^2}}{v}\nonumber\\
 U^i_j &=&  - \beta_t u_{ij}
 \label{DiracPosition1_1}
\end{eqnarray}

\section{Conclusions and discussion}

\label{SectCon}

In the present paper we expressed the emergent gauge field and emergent vierbein through the tensor of elastic deformations in the given model of Weyl semimetal. As follows from the general theory \cite{Z2015,VZ2014NPB}, the spin connection does not appear in the low energy effective field model of the fermionic quasiparticles. Therefore, we deal with a kind of emergent gravity with torsion, in which the emergent $SO(3,1)$ curvature vanishes. This is the case of the so - called teleparallel gravity (see \cite{VolovikZubkov2014} and references therein).

We follow the prescription for the modification of hopping parameters in the presence of elastic deformations proposed in \cite{WeylTightBinding,Vozmediano}. The calculated expressions may be used for qualitative investigation of various effects in Weyl semimetals. In particular, the complete expression for the Stodolsky effect may be given. (This is the correction to Aharonov - Bohm effect due to torsion. In \cite{Z2015} the topological contribution to this effect was given, while the total expression should contain the contributions proportional to the Gruneisen parameters $\beta_t,\beta_t^\prime,\beta_r$.) Besides, having in hand the expression for the emergent vierbein we are able to calculate the Nieh - Yan term \cite{Parrikar2014}, that, possibly enters expressions for the chiral anomaly in Weyl semimetals. Although we started from the model that does not refer directly to any particular crystal, we do not exclude, that our results may be relevant for the description of real materials in a certain approximation. This would open the possibility to check experimentally the appearance of Nieh - Yan term in the chiral anomaly, and, thus to resolve the ambiguity in the theoretical expression for chiral anomaly in the presence of gravity with torsion mentioned in the Introduction.

The most promising development of the present study may be related to the similar consideration of the tight - binding models that describe real Weyl and Dirac semimetals (TaAs, Cd$_3$As$_2$, Na$_3$Bi). Those models should be more complicated because of the non - cubic crystal structure. Therefore, the direct extention of the method developed in the present paper to the consideration of these materials should be complicated technically (though obvious theoretically).

M.A.Z. is greatful to Laboratoire de Math\'ematiques et Physique Th\'eorique (Tours, France), where the work was initiated,   for kind hospitality, and kindly acknowledges useful discussions with M.N.Chernodub and G.E.Volovik, as well as the private communications with M.Vozmediano. A.C. acknowledges the Spanish MECD Grant No. FIS2011-23713, the European Union structural funds and the Comunidad de Madrid MAD2D-CM Program (S2013/MIT-3007), and the European Union Seventh Framework Programme under grant agreement no. 604391 Graphene Flagship for financial support.
The  work of  M.A.Z. was supported by Far Eastern Federal University, by Ministry of science and education of Russian Federation under the contract 02.A03.21.0003, and by grant RFBR 14-02-01261.

\end{document}